\documentclass[showpacs,floatfix,amsmath,amssymb,prb]{revtex4}
\usepackage[dvips]{graphicx}
\usepackage{latexsym}
\usepackage{graphicx}
\usepackage{times,psfrag,subfigure}
\usepackage{amsmath}
\usepackage{dcolumn}
\newcommand{\be}{\begin{equation}}
\newcommand{\ee}{\end{equation}}
\newcommand{\ba}{\begin{eqnarray}}
\newcommand{\ea}{\end{eqnarray}}
\newcommand{\baa}{\begin{eqnarray*}}
\newcommand{\eaa}{\end{eqnarray*}}

\begin{document}

\title{Spin Precession and Oscillations in Mesoscopic Systems}
\author{Martin Y. Veillette}
\email{veillet@physics.ucsb.edu}
\author{Cristina Bena}
\email{cristina@physics.ucsb.edu}
\author{Leon Balents}
\email{balents@physics.ucsb.edu}
\affiliation{Department of Physics,
University of California at Santa Barbara,  Santa Barbara, CA-93106}
\date{\today}

\begin{abstract}
  We compare and contrast magneto-transport oscillations in the fully
  quantum (single-electron coherent) and classical limits for a simple
  but illustrative model.  In particular, we study the induced
  magnetization and spin current in a two-terminal double-barrier
  structure with an applied Zeeman field between the barriers and spin
  disequilibrium in the contacts.  Classically, the spin current shows
  strong tunneling resonances due to spin precession in the region
  between the two barriers.  However, these oscillations are
  distinguishable from those in the fully coherent case, for which a
  proper treatment of the electron phase is required.  We explain the
  differences in terms of the presence or absence of coherent multiple
  wave reflections.
\end{abstract}
\maketitle

\vspace{0.15cm}



\section{Introduction}
The recent progress in controlling and manipulating spins in
semiconductors has attracted considerable interest in the prospect of
developing electronics based on spin degrees of freedom\cite{Datta,
  Awschalom}.  Numerous recent magneto-transport experiments expose
``spin coherence'' in the sense of long-lived oscillatory phenomena
associated with spin precession\cite{Kikkawa0, Kikkawa1}.  Some such oscillations, however, are
expected to persist even in the classical limit as a consequence only
of (approximate) spin-rotational invariance.  The aim of this paper is
to compare and contrast spin transport in simple models of classical
and truly single-electron phase coherent conductors.  We also give
a set of general hydrodynamic transport equations that apply in the
classical regime -- and can be applied for quantitative modeling of a
variety of experimental geometries. 

Theoretically, although the electron spin is a quantum object at
heart, its quantum nature has been largely ignored in studies of
semiconductor spin transport.  Most approaches are based on a
semi-classical Boltzmann formalism, in which quantum mechanical effects
appear only in the calculation of scattering rates and Fermi
occupation factors\cite{Valet, Johnson1}.  The resulting semi-classical transport
 equations were successfully applied in analyzing spin
injection into semiconductors and in studying spin-polarized transport
in inhomogeneous doped semiconductors \cite{Johnson, Jedema}.  This
approach essentially ignores all quantum interference effects, and is
expected to hold when the inelastic mean free path is sufficiently
short -- i.e. for not too low temperature and for large samples.
However, experiments on mesoscopic conductors allow a detailed
exploration of phase coherent phenomena, as has been amply
demonstrated for charge transport.\cite{DattaBook}  

To search for single-electron spin coherence, it is crucial to have
some idea of its distinguishing features.\cite{ImryBook} The search is complicated
by the existence of essentially {\sl classical} spin oscillations.  In
particular, in any system which, in absence of applied fields, can be
approximated as spin-rotationally invariant, a continuity equation
(modified by a precessional ``mode-coupling'' term) exists:
\be
\partial_t S_\mu + \partial_\nu I_{\mu\nu}= g \mu_B
 \epsilon_{\mu\nu\lambda} S_\nu B_\lambda
\label{eq:continuity}
\ee 
where $S_\mu$ is the spin density, $I_{\mu\nu}$ is the spin
current tensor (current of the $\mu^{\rm th}$ component of spin
running along the $\nu^{\rm th}$ direction).  Here $g$ and $\mu_B$ are
the gyromagnetic factor and Bohr magneton. In a single-channel
conductor, this becomes 
\be \partial_t \vec{S} + \partial_x \vec{I}= g
\mu_B (\vec{S} \times \vec{B}),
\label{eom1}
\ee 
where $\vec{S}$ and $\vec{I}$ are the spin and spin-density current.  These
two equations are operator identities, and hence trivially hold
for the expectation values of the spin current and density, since they
are linear.  No assumption whatsoever is made regarding
single-electron phase coherence, and indeed these would hold even in a
non-degenerate electron gas at high temperature with large amounts of
inelastic scattering, provided SU(2) spin-rotation symmetry is only
weakly broken.  Clearly then, spin precession -- and associated
magnetotransport oscillations -- may occur even in strongly
incoherent (indeed classical) media.  

To elucidate the aspects of magnetotransport oscillations which {\sl
  do} rely on single electron coherence, we consider a simple
situation in which the classical and quantum limits can be compared
and contrasted.  We focus on a double barrier structure consisting of
two tunneling barriers separated by a region where a magnetic field is
applied perpendicularly to the system (See Fig.\ref{DoubleBarrier}).
Incidentally, such Zeeman fields tailored on a mesoscopic scale have
been ingeniously achieved by making use of the Overhauser effect
through which optically polarized spin nuclei induce a Zeeman magnetic
field on the conduction electrons through the contact interaction.  In
this way magnetic fields of the order of 1 Tesla extending over a
range of microns have been obtained \cite{Kikkawa2}.  We describe (by
definition) the coherent quantum limit of this system using the
Landauer-B\"uttiker formalism for electron wave scattering using the
S-matrices for the two barriers.  The classical limit is more delicate
to define.  Indeed, varying degrees of decoherence can occur via
different physical mechanisms, by which individual electrons
interchange energy between each other or with other degrees of freedom
(e.g. phonons).  The exchange of energy leads to dephasing -- via
$d\phi/dt = E/\hbar$ -- as well as energy equilibration.  To maximize
the contrast with the coherent case, we consider the most extreme
classical limit, suggested by B\"uttiker\cite{Buttiker1986, Knittle}, in which
local electronic reservoirs are {\em attached} to the wire (see Fig.
\ref{DoubleBarrierc}).  These local equilibrium regions can be
considered either as fictitious or real voltage probes.  The
randomization of the phase occurs by removing electrons from the phase
coherent channels and reinjecting them without any phase
relationships.  Because in this approach electrons are actually interchanged between  the reservoirs and the wire,  not only is the single
electron phase randomized but also energy and particle
number are locally equilibrated.  Hence, this picture can be viewed as a discrete
limit of a hydrodynamic approach: while in hydrodynamics there is
continuous local equilibrium at each point, in this model local
equilibrium is realized only at the two points of the voltage probes.

>From our calculations we find there are qualitative differences between incoherent and coherent propagation. We demonstrate that for phase coherent electrons, spin (charge) current can be generated from charge (spin) chemical potential, a situation which has no classical analog. These results can be understood in terms of quantum interference between multiple partial waves. Constructive interference between multiple reflections is determined by the added phase due to the applied magnetic field. We find that this can be put to use to create a spin valve where small changes in magnetic field lead to large changes in conductance. In the classical regime, the phase loss induced by the extra contacts obliterate the quantum interference but nevertheless spin precession arises. However, it is nature is different and is to be interpreted as arising from resonances instead of quantum interferences.

The remainder of the paper is organized as follows.  In Sec.~II. we give a description of spin transport in the Buttiker-Landauer model and derive the magnetoconductance of a 1-d wire for coherent electron propagation. In Sec.~III. we consider incoherent electron propagation and determine the spin hydrodynamic equations in the ballistic regime . The results of the two approaches are elaborated and discussed in Sec.~IV.  Finally, in Sec.~V. we present possible experimental implementations and a summary.

\section{Model: Scattering Approach}

\begin{figure}[ht]
\begin{center}
\includegraphics[width=3.0in]{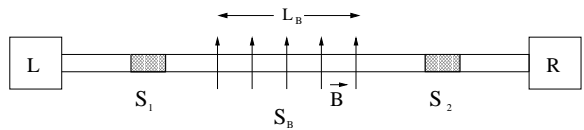}
\end{center}
\vspace{0.05in}
\caption{Sketch of the double barrier. Spin-polarized currents are
injected in the wire through the left (L) and right (R) contacts.
The two potential barriers are characterized by scattering matrices $S_1$ and
$S_2$ whereas the
the mid-region is subject to a tunable perpendicular magnetic field whose
effect can be represented by the scattering matrix $S_B$.}
\label{DoubleBarrier}
\end{figure}

We adopt a scattering approach to treat coherent electron propagation
through a double barrier, focusing for simplicity on the ballistic
regime in which the elastic mean free path $l_e$ is much larger than
the channel length $L_c$.  The corresponding formalism is based on the
scattering matrix (S-matrix), which connects incoming
$\hat{a}$ and outgoing $\hat{b}$ electron annihilation operators in
the (L) left and (R) right leads, such that 
\be
\begin{pmatrix}
\hat{b}_L \\
\hat{b}_R
\end{pmatrix}
 =
\begin{pmatrix}
 S_{LL} & S_{LR}\\
 S_{RL} & S_{RR}
\end{pmatrix}
\begin{pmatrix}
\hat{a}_L \\
\hat{a}_R
\end{pmatrix}.  \ee 
Here, we consider only two allowed propagation channels (one for each
spin), so that each $S_{i j}$ is a $2 \times 2$ spin matrix. The
corresponding creation operators $\hat{a}^\dagger$ and
$\hat{b}^\dagger$ are similarly connected by the hermitian conjugated
S-matrix.

In the Landauer-B\"uttiker approach, the incident flow of electrons into
the wire is determined by the electron distribution in the reservoirs.
We consider paramagnetic contacts spin polarized along a given
direction $\hat{n}$ such that each spin band is characterized by its
own Fermi-Dirac distribution \cite{Tang}.  In the proper spin basis,
the occupation numbers in the contacts are given by 
\be \langle
a^{\dagger}_{m i \alpha} a^{\phantom{\dagger}}_{n j \beta} \rangle =
\delta_{m n} \delta_{i j} \delta_{\alpha \beta} f_\alpha(\epsilon_{ m
  i \alpha}),
\label{distribution}
\ee where the Fermi distribution $f_{\alpha}$ is \be
f_\alpha(\epsilon)= \frac{1}{e^{(\epsilon-\mu_{\alpha})/k_BT}+1}.  \ee
Here $a^{\dagger}_{m i \beta}$ and $a^{\phantom{\dagger}}_{m i \beta}$
are respectively creation and annihilation electron operators in the
contact $i=L/R$, with spin $\alpha=\uparrow/\downarrow$ and quantum
number $m$ representing other degrees of freedom besides the spin. The
energy $\epsilon$ is the single particle excitation energy. This
double Fermi distribution function is realized provided that the
interactions in the contacts preserve spin, but randomize the quantum
number $m$ (e.g. the kinetic momentum). Thus the double Fermi
distribution function describes two distinct (spin up and down)
populations of electrons whose numbers are separately conserved but
which may exchange energy and hence share the same temperature.  We
define a spin chemical potential $\vec{\mu}$ to be $\left(\mu^{\uparrow} -\mu^{\downarrow} \right) \hat{n}$
where $\hat{n}$ is the direction of the spin polarization, and the
charge chemical potential as $\mu=(\mu^{\uparrow} +
\mu^{\downarrow})/2$.  

The spin chemical potential determines the occupations (partially
traced density matrix) 
\ba \sum_{m} \langle a^{\dagger}_{m i \alpha}
a^{\phantom{\dagger}}_{m i \beta} \rangle = \delta_{\alpha \delta} N/2 + N(0) \left(\mu^0_i
  \delta_{\alpha \beta} + \frac{1}{2} \vec{\mu}_i \cdot
  \vec{\sigma}_{\beta \alpha}\right) , \ea 
where $\mu^0=\mu-\epsilon_F$ is the  charge chemical potential deviation due to electrostatic potentials, $\vec{\sigma}=(\sigma^x, \sigma^y, \sigma^z)$ are the Pauli matrices
and $N$, $N(0)$ are the number of particles and the density of states per spin at the Fermi energy respectively. Note that  the spin polarization is related to the spin chemical
potential via $\vec{S}= N(0) \vec{\mu}/2$.  We emphasize that this is a different situation than the
case of ferromagnetic contacts, in which the majority and minority carriers share the same chemical: the spin chemical potenital vanishes and  the spin polarization is due to a non-zero internal
exchange field.  Experimentally, a non-vanishing spin chemical potential
can be realized in electron-doped III-V semiconductors by optical methods
which highly polarize the doped conduction band \cite{Malajovich}.  In
the absence of magnetic impurities and spin orbit interaction, the
spin polarization is a conserved quantity and a double Fermi
distribution is achieved.  Hence, this distribution can be
realized for times larger than the inelastic scattering time, to allow
for thermalization of each spin specie, but smaller than the spin-flip
scattering time in order to preserve the spin polarization. 

According to Landauer, the charge $I^0$ and spin $\vec{I}$ currents in
the lead $i$ can be written as  
\ba I^0_i&=&e
\sum_{\alpha} \sum_{m} v_{m i} \left( \langle a_{m i \alpha}^{\dagger} a_{m i
    \alpha} \rangle- \langle b_{m i \alpha}^{\dagger}
  b_{m i \alpha} \rangle \right)/L , \\
\vec{I}_i&=&\frac{\hbar}{2} \sum_{\alpha \beta} \sum_{m} v_{m i} \left( \langle
  a_{m i \alpha}^{\dagger} \vec{\sigma}_{\alpha \beta} a_{m i \beta}
  \rangle- \langle b_{m i \alpha}^{\dagger}\vec{\sigma}_{\alpha \beta}
  b_{m i \beta} \rangle \right)/L, \ea 
where $v_{m i}$ is the velocity of the level $m$ in the lead $i$ along the $x$-direction and $L$ is the lenght of the wire. Note from hereon we will make use of the fundamental units where $\hbar=1$ and $e=1$. We use the convention that the incoming currents
are positive.  For small chemical potential (relative to the Fermi energy), we can consider the scattering matrix to be energy-independent. In the zero-temperature limit, we arrive to the current equations  
\ba
I_i^{0}&=&\frac{1}{2\pi}\left\{2\mu_i^0 - \sum_{j=R,L}
\left(\mu_j^0 Tr[S^{\dagger}_{ji} S_{ij}]+
\frac{\vec{\mu}_j}{2}\cdot Tr[ \vec{\sigma} S^{\dagger}_{ji}
S_{ij}]\right) \right\},  \nonumber \\
\vec{I}_i&=&\frac{1}{4 \pi}\left\{ \vec{\mu}_i
- \sum_{j=R,L} \left(\mu_j^0 Tr[S^{\dagger}_{ji} \vec{\sigma}
S_{ij}] +
\sum_{\nu}\frac{\mu_j^{\nu}}{2} Tr[\sigma^{\nu} S^{\dagger}_{ji}
\vec{\sigma} S_{ij}] \right) \right\}.
\label{current}
\ea

We note that the spin (charge) chemical potential can also generate
charge (spin) current. Also one can readily check
that these equations are gauge invariant, i.e. a uniform shift of the
charge chemical potentials does not generate any currents.  We also
emphasize that due to the mode coupling term in Eq. \ref{eq:continuity}, the spin current need not be the same in both leads. This is contrast to the charge current where charge conservation implies that $I^0_L= -I^0_R$.

To derive an explicit expression for
the entire scattering matrix $S$ of the system, we first describe the effect
of the potential barriers $S_1$ and $S_2$ sketched in Fig.
\ref{DoubleBarrier} by using the most general spin-independent scattering
matrices for a two-channel system,
\be
S_{j}= e^{i \eta_j} \left[ i e^{ i \xi_j \tau_z} \sin (\gamma_j) +
\tau^x \cos (\gamma_j) \right] \otimes \sigma^0,
\label{Sj}
\ee
where $\tau^0,\vec{\tau}$ are the $2 \times 2$ identity and
Pauli matrices respectively acting in the basis of incoming and outgoing
electrons. Each barrier transmits and
reflects electrons with probability $T_j=\cos^2(\gamma_j)$
and $R_j=\sin^2(\gamma_j)$, respectively. The overall phase $\eta_j$
is merely a propagating phase for all processes and without loss of
generality will be set to zero from now on. The phase $\xi_j$
introduces an asymmetric phase shift between the backscattering
of electrons coming from the left and the right.

We also ascribe a scattering
matrix to the spin precession generated by the magnetic field in
the double barrier region. For a Zeeman field pointing along the
$z$-direction, the spin-dependent phase winding yields
\be
S_{B}= \tau^x \otimes \exp (i \delta \sigma^0 - i \theta \sigma^z/2),
\ee
where $\delta=k_F L'$ and $\theta/2= g u_B B^z L_B/ 2 v_F$ are
respectively the quantum mechanical phase acquired by free electrons
propagating over an interbarrier distance $L'$, and the
spin-dependent phase due to the winding in the magnetic field
over a length $L_B$. Here
$g, u_B, B^z, v_F,k_F$ are respectively the gyromagnetic ratio,
the Bohr magneton, the magnetic field, the Fermi velocity and the Fermi
wavevector. Physically, the electron spins undergo a rotation by an angle 
$\theta$ in the $x-y$ plane upon crossing the magnetic field region.
The S-matrix of the entire system is obtained by considering the combined
effects of the potential barriers $S_{1}$, $S_{2}$ and the phase winding
$S_{B}$  and summing over all possible electron paths. Although we have
determined the full scattering matrix, for ease of presentation we will set
$\xi_1=\xi_2=0$ and $\gamma_1=\gamma_2=\gamma$ to obtain the total
S-matrix
\begin{eqnarray}
S=\frac{1}{\sigma^0+\sin^2(\gamma) e^{i (2 \delta \sigma^0 -
\theta \sigma^z)}}
\big\{i \tau^0 \sin(\gamma) \otimes [\sigma^0+
e^{i (2 \delta \sigma^0 - \theta \sigma^z)}]+
\tau^x \cos^2(\gamma)\otimes
e^{i(\delta \sigma^0- 1/2 \theta \sigma^z)}\big\}
\label{smatrix}.
\end{eqnarray}


The denominator is a result of multiple reflections between the two
barriers. Notice that $S$ is periodic with respect to $\theta$ with a
period of $4 \pi$ which corresponds to two complete spin rotations.
Substituting the scattering matrix
into Eq. (\ref{current}), we obtain the charge and spin currents
in the two leads:
\begin{eqnarray}
&&I_L^0=\frac{T^2}{2\pi}\left( \frac{1}{|\Delta_1|^2}+
\frac{1}{|\Delta_2|^2}\right) ( \mu_L^0 -\mu_R^0) +
\frac{T^2}{4\pi}\left( \frac{1}{|\Delta_1|^2}-
\frac{1}{|\Delta_2|^2}\right) ( \mu_L^z -\mu_R^z)
\label{currentq1} , \nonumber \\
&&I_L^{x}=\frac{1}{4 \pi}
\left[ \mu_L^x-4 ( 1-T) \cos( \frac{\theta}{2}-\delta) \cos(
\frac{\theta}{2}+\delta)Re\left( \frac{\mu_L^x+i \mu_L^y}{\Delta_1^* \Delta_2}\right) -
T^2 Re\left( \frac{\mu_R^x+i \mu_R^y}{\Delta_1^* \Delta_2}\right)  \right]
\label{currentq2}, \nonumber \\
&&I^y_L=\frac{1}{4 \pi}
\left[ \mu_L^y-
4 ( 1-T) \cos( \frac{\theta}{2}-\delta) \cos( \frac{\theta}{2}+\delta)
Im\left( \frac{\mu_L^x+i \mu_L^y}{\Delta_1^* \Delta_2}\right)  -
T^2 Im\left( \frac{\mu_R^x+i \mu_R^y}{\Delta_1^* \Delta_2}\right)  \right]
\label{currentq3} , \nonumber \\
&&I_L^z=\frac{T^2}{8 \pi}\left( \frac{1}{|\Delta_1|^2}+
\frac{1}{|\Delta_2|^2}\right) ( \mu_L^z -\mu_R^z) +
\frac{T^2}{4\pi}\left( \frac{1}{|\Delta_1|^2}-
\frac{1}{|\Delta_2|^2}\right) ( \mu_L^0 -\mu_R^0),
\label{currentq4}
\end{eqnarray}
where $\Delta_{1/2}=\exp( -i \delta \pm i \theta/2) +( 1-T)
\exp( i \delta \mp i \theta/2)$. The currents in the right lead can be
simply obtained by performing the transformation $R \leftrightarrow L$.
>From the current formula, we can read off that the spin currents in
the $x-y$ plane ($I^x,I^y$) depend only on the in-plane spin chemical
potentials $\mu^x, \mu^y$, whereas the charge current and the spin
current $I^0$ and $I^z$ depend on $\mu^0, \mu^z$. This result can be
traced back to the fact that the scattering matrix does not have off
diagonal elements in the spin basis, so that the spin
up and down currents in the $z$-basis are separately preserved. Since
the currents $I^0$ and $I^z$ are respectively a sum and a difference
of these currents, this explains why $I^0$ and $I^z$ are proportional to
the difference of chemical potentials in each lead. It is interesting
to notice that it is possible to drive a spin(charge) current by an
electrostatic(spin) chemical potential.   This situation has no
classical analog, as we will see below.

\section{Classical Propagation}
\label{sec:class-prop}
\begin{figure}[ht]
\begin{center}
\includegraphics[width=3.0in]{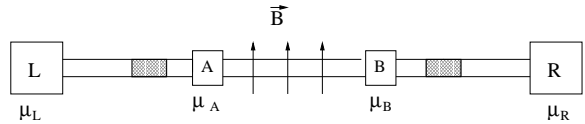}
\end{center}
\vspace{0.05in}
\caption{Sketch of the double barrier system. Spin chemical potential is
applied on the left (L) and right (R) contacts. The two voltage probes $A$ and $B$ randomize the phase of the injected electrons
and destroy the effects of quantum coherence}
\label{DoubleBarrierc}
\end{figure}

In general, we expect differences from the above results, derived
assuming coherent electron propagation, when a single electron's phase
is significantly scrambled in the time required to propagate between
the two barriers.  As discussed in the introduction, in this
section we consider an extreme classical limit, in which not only
is this single electron phase coherence completely lost, but also the
electrons are allowed to locally equilibrate inside the structure, at
the locations of the voltage probes A and B (with chemical potentials
$\mu_A,\mu_B$) in
Fig.~[\ref{DoubleBarrierc}].  Such local equilibrium is in general
required to achieve fully classical behavior, e.g. Ohm's law for
charge conduction.

Assuming the effect of scattering by the barriers can be modeled
classically (on distances longer than the length of the barriers but
smaller than the inter-barrier distance), we may determine the spin
current using the continuity equation,Eq.~(\ref{eom1}), and the
classical hydrodynamic equation,
\begin{eqnarray}
\partial_t \vec{I}+ v_F^2 \partial_x \vec{S}&& =
-g \mu_B (\vec{I} \times \vec{B}) - \vec{I}/\tau^s_{\rm el}
\label{eom}
\end{eqnarray}
where $\tau^s_{\rm el}$ is the relaxation time for spin currents, which
we assume to be governed by the elastic collision time.  We assume
$\tau^s_{\rm el} \neq 0$ only within the regions of the barriers.  Then 
in the steady state, Eq.~(\ref{eom1}) implies $\vec{I}$ is constant
through each barrier, and integrating Eq.~(\ref{eom}) through the
barrier regions gives
\begin{eqnarray}
  \vec{I}_L=\frac{v_F \tau^s_{\rm el}}{2 \pi L_{\rm
      bar}}(\vec{\mu}_L-\vec{\mu}_A) 
  \nonumber ,\\
  \vec{I}_R=\frac{v_F \tau^s_{\rm el}}{2 \pi L_{\rm
      bar}}(\vec{\mu}_R-\vec{\mu}_B), 
\end{eqnarray}
where $L_{\rm bar}$ is the length of the contact. Comparing with the Landauer-B\"uttiker result 
for the spin conductance of a single barrier, we will take
$\frac{v_F \tau^s_{\rm el}}{2 \pi L_{\rm bar}} = T/4\pi$ in the
following. Similarly solving Eqs.~(\ref{eom1},\ref{eom}) in the region
between A and B gives
\begin{eqnarray}
  \vec{I}_L=
  \frac{1}{4\pi}(\vec{\mu}_A-\hat{R}_{\theta}\vec{\mu}_B)
  \nonumber ,\\
  \vec{I}_R=\frac{1}{4\pi}
  (\vec{\mu}_B-\hat{R}_{\theta}\vec{\mu}_A),
\end{eqnarray}
where, in the $(x,y,z)$ basis,
$\hat{R}_{\theta}$ reads\be
\hat{R}_{\theta}=
\begin{pmatrix}
\cos(\theta)& -\sin(\theta)&0\\
\sin(\theta)&\cos(\theta)&0\\
0&0&1
\end{pmatrix}.
\ee\

The similar but more trivial classical result for the charge currents
is 
\begin{eqnarray}
I_L^0=\frac{T}{\pi}(\mu^0_L-\mu^0_A)=\frac{1}{\pi}(\mu^0_A-
\mu^0_B) \nonumber , \\
I_R^0=\frac{T}{\pi}(\mu^0_R-\mu^0_B)=\frac{1}{\pi}(\mu^0_B-
\mu^0_A)  .
\end{eqnarray}

Solving these Kirchoff's-like equations for the spin chemical potentials
$\vec{\mu}_A$ and 
$\vec{\mu}_B$ of the reservoirs $A$ and $B$, we find
\ba
\begin{pmatrix}\vec{\mu}_A\\ \vec{\mu}_B \end{pmatrix}
&=&\hat{K}
\begin{pmatrix} 1+T&\hat{R}_{ \theta}\\\hat{R}_{\theta} & 1+T
\end{pmatrix}\begin{pmatrix}\vec{\mu}_L\\ \vec{\mu}_R\end{pmatrix}
\label{potentials}
\ea
as well as the
currents:
\ba
I^0_L&=& - I^0_R =\frac{1}{\pi} \frac{T}{2+T}
(\mu^0_L-\mu^0_R)
\label{currentc1}\\
\begin{pmatrix}\vec{I}_L\\ \vec{I}_R \end{pmatrix}
&=&\frac{1}{4\pi} \left[ T \hat{K}
\begin{pmatrix} 
1+T &\hat{R}_{ \theta} \\ \hat{R}_{\theta} & 1+T
\end{pmatrix}
\begin{pmatrix}\vec{\mu_L}\\ \vec{\mu}_R
\end{pmatrix}+
\begin{pmatrix} 
T&0\\0&T\end{pmatrix}
\begin{pmatrix}
\vec{\mu_L}\\ \vec{\mu}_R
\end{pmatrix} 
\right] .
\label{currentc2}
\ea

Here we defined
\begin{eqnarray}
\hat{K}&&=
T [\hat{R}^2_{\theta}-{\hat{I}} (1+T)^2]^{-1}
\nonumber\\&&=
\frac{T}
{1+(1+T)^4-2(1+T)^2 \cos(2 \theta)}
\begin{pmatrix}
\cos(2\theta)-(1+T)^2&
\sin(2\theta)&0\\-\sin(2\theta)&\cos(2\theta)-(1+T)^2&0\\
0&0&-\frac{1+(1+T)^4-2(1+T)^2 \cos(2 \theta)}{T(2+T)}
\end{pmatrix}
\end{eqnarray}
and $\hat{I}$ is the $3\times3$ identity matrix. For weak tunneling
($T \ll 1)$, the matrix $\hat{K}$ is responsible for the resonances in
the current function.  Note that in addition to
dephasing, each reservoir randomized the electronic momentum and
generated an additional resistance.  This extra resistance can be read
off from the charge current where the total transmission trough the
system is given by the fraction $\frac{T}{2+T}<1$, to be compared to
the result for coherent electrons $\frac{T}{2-T}$. The former can be
written as $(\frac{2R}{T} +1)^{-1}$ whereas the latter is
$(\frac{2R}{T} +1+2)^{-1}$. For coherent propagation the resistance is
the sum of $\frac{2R}{T}$ corresponding to the intrinsic resistance of
the potential barriers and $1$, the quantum resistance associated to
the left and right contacts.  In the case of incoherent propagation,
there is an additional factor of $2$ due to the quantum resistances of
the two voltage probes.

\section{Discussion}

\subsection{Low transmission limit}

\begin{figure}[ht]
\begin{center}
\includegraphics[width=3.0in]{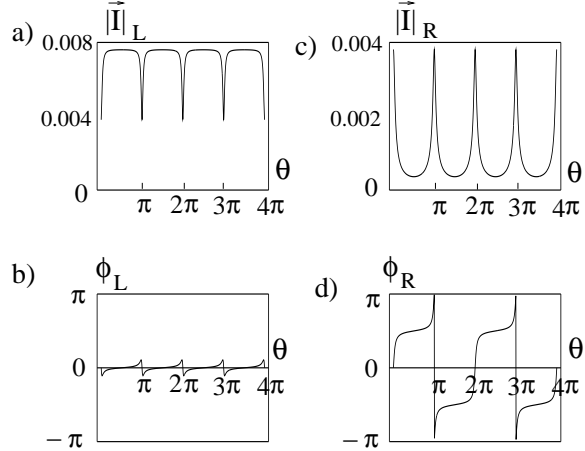}
\end{center}
\vspace{0.10in}
\caption{Incoherent propagation for opaque barriers:
The magnitude and the
direction in the $xy$ plane of the spin current in the left (L) and right
 (R) leads respectively as a function of $\theta$. A spin chemical potential
 $\mu^x_{L}=1$ is applied on the left lead. The transmission $T=0.1$ }
\label{muzq}
\end{figure}

Phase coherence is known to affect transport properties in the charge
channel. However, much less is known in the spin sector. One may
expect that these differences, if present, to be enhanced in the low
transmission limit where multiple reflections in the intrabarrier region
occurs. In this problem, we first point out that in the case of incoherent
propagation spin and charge are completely decoupled, i.e., classically the spin chemical potential
cannot give rise to charge current and vice versa. The charge conductance
and spin conductance along the z-axis are given by $G_c=\frac{1}{\pi}
\frac{T}{2+T}$ and $G^z_s=\frac{1}{4 \pi} \frac{T}{2+T}$
respectively. In the geometry investigated, the Zeeman magnetic field
does not exert any force on the $z$ component of the spin or on the electric charge, therefore
the spin and charge  are separate conserved quantities and each has an
associated conductance that is independent of the applied magnetic field.
However, for coherent electron propagation, spin and charge mix as
shown in Eq. (\ref{currentq4}). This effect is due to the filtering action of the magnetic field. In the low
transmission limit the tunneling transmission is strongly
enhanced  for specific values of the applied magnetic field (see
Fig.\ref{muzq}). This can be explained on the basis of constructive
interference between multiple electron wave reflections at the scattering
barriers. Analyzing Eq. (\ref{smatrix}), this condition translates for an
electron with spin $\uparrow/\downarrow= +/-$
as $( 2\delta \pm \theta + \pi= 2 \pi n)$. Hence, the spin
current is periodic with respect to $\theta$ with a period of $2 \pi$, and
the different peaks corresponds to constructive interference for spin up
and down electrons. In the limit $T \rightarrow 0$,
the charge and spin currents at the resonant values are
\begin{eqnarray}
I_L^0&&=\frac{1}{2 \pi}(\mu_L^0-\mu_R^0)\pm
\frac{1}{4 \pi}(\mu_L^z-\mu_R^z),
\nonumber \\
I_L^z&&=\frac{1}{8 \pi}(\mu_L^z-\mu_R^z)\pm
\frac{1}{4 \pi}(\mu_L^0-\mu_R^0),
\end{eqnarray}
while the currents in the right lead are obtained by interchanging
$L$ and $R$. 

For simplicity in Fig.\ref{muzq} we depict only the currents
in the left lead $I_L^0,I_L^z$ in the case of the only non-zero applied
chemical potential $\mu_L^z-\mu_R^z=1$. In this geometry, a double barrier system can be seen as a spin valve where a small deviation in the magnetic field from the constructive interference conditon lead to a large change in conductance.  Also we notice that, since the charge and the spin are carried by the
 same particle, a positive (negative) charge current is also generated
when a spin up (down) is transmitted. Hence, this phenomenon is essentially
linked to the particle/wave duality of the electron.
\begin{figure}[ht]
\begin{center}
\includegraphics[width=2.0in]{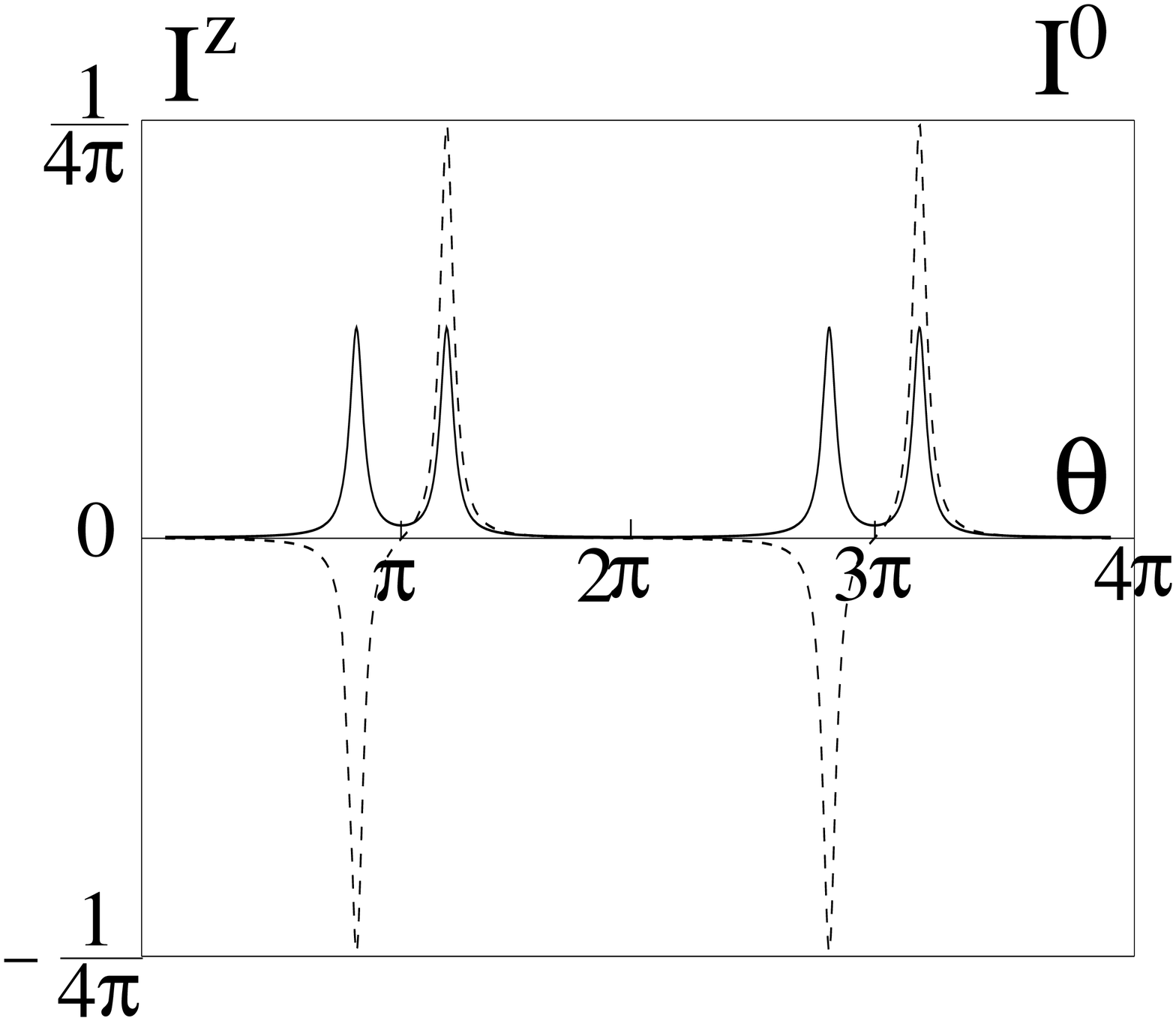}
\end{center}
\vspace{0.10in}
\caption{Charge current $I^0$ in dash line and $z$-spin current $I^z$
shown by the solid curve for fully coherent tunneling. A spin chemical
potential $\mu^z_L-\mu^z_R=1$ is applied, leading to a non-zero charge current.
We took $T=0.1$ and $\delta=0.3$.}
\vspace{0.05in}
\label{muxq}
\end{figure}

\begin{figure}[ht]
\begin{center}
\includegraphics[width=3.0in]{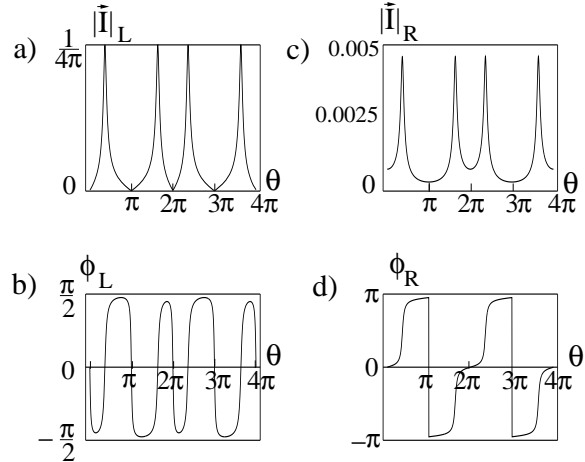}
\end{center}
\vspace{0.10in}
\caption{Coherent propagation for opaque barriers: The magnitude and the
direction in the $xy$ plane
of the spin current in the left and right leads
respectively as a function of $\theta$. The transmission $T=0.1$ and
$\mu^x_{L}=1$.}
\vspace{0.05in}
\label{muxc}
\end{figure}

To further illustrate and contrast the results of Eqs.
(\ref{currentq1}) and
(\ref{currentc1}-\ref{currentc2}), we present results for the spin current
in the presence of
spin chemical potentials along the $x$ or $y$ axis.
Classically, the spin would merely
precess in the $x-y$ plane due to the magnetic field. A strong transmission
occurs when spins perform a complete revolution
upon a closed trajectory between the barriers, i.e. the acquired phase
$2 \theta = 2 n \pi$, and
the partial spin reflections
are added in phase giving rise to a spin resonance.
If the spins precess
by  $\theta=2 k \pi$ respectively $(2 k+1) \pi$
while traveling from one potential barrier to the other
the transmitted spin current
in the right lead points along the positive, respectively negative $x$
direction as depicted in Fig. \ref{muxc}.
The value of the currents at such resonances are
\begin{eqnarray}
I_L^x=\frac{1}{4 \pi} \frac{T}{2+T} (\mu_L^x\pm\mu_R^x)
\\
I_L^y=\frac{1}{4 \pi} \frac{T}{2+T} (\mu_L^y\pm\mu_R^y)
\end{eqnarray}
where the $\pm$ signs correspond to resonances at $\theta= 2k \pi , (2
k +1)\pi$ respectively and the currents in the right lead are
determined as usual by interchanging the $L$ and $R$ symbols.  Far off
resonance, the spins are added out of phase and a small current of
order $T^2$ is generated.  On the other hand, for fully coherent
electrons, the current peaks do not occur upon full spin revolutions
in the barrier but, as mentioned earlier, are determined by the
constructive interference between multiple partial waves.
Interestingly, as shown in Fig. \ref{muxq}, we point out that the spin
current in the left lead points along the $y$ axis for broad ranges of
the applied magnetic field.  It is also straightforward to show that
in this case (i.e. $\mu^x_L \neq 0$), at the points of constructive
interference, the ratio of spin currents in the right lead is
\begin{equation}
\frac{I_R^x}{I_R^y}=\frac{1-R}{1+R} \cot (\theta).
\end{equation}
Hence in the low transmission limit, the spin currents emerge with a
spin direction mostly along the $y$ axis. It is worth emphasizing that
even if one considers all phase factors in the potential barriers
(Eq.~(\ref{Sj})) the result remains unchanged and is consequently a robust
prediction.


\subsection{High transmission limit}

Most of our discussion is focused on the limit of low transmission of
the barriers when the most important effects are due to resonant
tunneling through the two barriers. Hence, in the high transmission limit,
electrons impinging on the potential barriers are most likely to be
transmitted, therefore one expects the interference effect resulting
from the multiple reflections to be reduced. Restricting our attention
to perfectly transmitting barriers ($T=1$), we find the currents in
the phase-coherent regime to be given by

\begin{eqnarray}
\vec{I}_L=\frac{1}{4\pi}(\vec{\mu}_L-\hat{R}_{\theta}\vec{\mu}_R)
\nonumber. 
\end{eqnarray}

Hence the currents are precessing in the magnetic field following
Eqs.~(\ref{eom}) as one might have expected.  However, this is to be
contrasted with the behavior in the phase incoherent regime, \ba
\begin{pmatrix}
I^x_L \\
I^y_L \\
I^z_L 
\end{pmatrix}
&=& 
\frac{1}{4 \pi} \frac{1}{8 \cos ( 2 \theta )-17}  
\begin{pmatrix}
-9+6 \cos (2 \theta) & -2 \sin (2 \theta) & 0 \\
2 \sin ( 2 \theta ) & -9 +6 \cos ( 2 \theta) & 0 \\
0 & 0 & \left( 8 \cos ( 2 \theta) -17 \right)/3 
\end{pmatrix}
\begin{pmatrix}
\mu^x_L \\
\mu^y_L \\
\mu^z_L 
\end{pmatrix} \\
&+&
\frac{1}{4 \pi} \frac{1}{8 \cos ( 2 \theta )-17}   
\begin{pmatrix}
3 \cos ( \theta) & -5 \sin ( \theta) & 0 \\
5 \sin ( \theta ) & 3 \cos ( \theta) & 0 \\
0 & 0 & -\left( 8 \cos ( 2 \theta) -17 \right)/3 \\
\end{pmatrix}
\begin{pmatrix}
\mu^x_R \\
\mu^y_R \\
\mu^z_R \\
\end{pmatrix}.
\ea The unexpected difference between the two previous results is
linked to the extra resistance introduce by the voltage probes $A$
and $B$, as discussed in section II. The voltage probes achieve phase
randomization by bringing in local equilibrium at the point of
contact. Although inelastic scattering is needed to implement
decoherence, the B\"uttiker dephasing model enforces the more stringent
condition of local equilibrium. The momentum randomization affect both
the transmitted and reflected current which show extra oscillations
with the magnetic field due to the backscattering of the electrons.




\section{Hydrodynamics, Experimental Implementation and Conclusions}

>From the above discussion, it is apparent that there are numerous
features distinguishing the classical precessional limit from truly
single electron phase coherent propagation.  Hence in analyzing
experiments, it is important to distinguish the various time scales
being measured, e.g. the phase coherence time $\tau_\phi$ beyond which
classical behavior obtains from $\tau_s$, beyond which magnetization
relaxes and all signs of precession are lost (precession can also be
destroyed experimentally by other means, e.g. inhomogeneous magnetic
fields characterized by $T_2$ times).  The identifying of the various
scattering times -- other than $\tau_\phi$ which is determined by the
loss of true interference phenomena -- should be made by their
definitions through hydrodynamic equations.  Such equations generalize
Eqs.~(\ref{eom1},\ref{eom}), whose utility is emphasized by the
simplicity of the classical calculations in Sec.~\ref{sec:class-prop}.
In hopes that such careful determinations of various time scales may
be made, we give the hydrodynamics appropriate for electrons in 
III-V semiconductors, including the possibly important effects of
Rashba interactions due to electron confinement.  For a one-dimensional wire, the equations yield 

\begin{eqnarray}
\partial_t \vec{S} +\partial_x \vec{I}&&=-\frac{\vec{S}}{\tau_s}+g \mu_B
 \vec{S} \times \vec{B}+\frac{\alpha_R m^*}{\hbar^2} (\vec{I} 
\times \hat{Q}),
\nonumber \\
\partial_t \vec{I}+ v_F^2 \partial_x \vec{S}&&=-\frac{\vec{I}}{\tau^s_{el}}
-\frac{e}{m^*} E_x \vec{S} -g \mu_B  \vec{I} \times \vec{B}
+\frac{ v_F^2 \alpha_R m^*}{\hbar^2} (\vec{S} 
\times \hat{Q})-\frac{g \mu_B \hbar n}{4 m^*}  \partial_x \vec{B} -\frac{u}{\hbar} \vec{S} \times \vec{I}
\end{eqnarray}
where $\alpha_R$ is the Rashba coefficient, $\hat{Q}$ is the direction
of the Rashba spin-splitting, $E_x$ is the bias electric field, $n$ is
the electron number density and $\tau_s$ is the spin relaxation time. The
last term with coefficient $u \sim e^2 k_F$ derived from the
electron-electron interaction yield non-linear equations which
severely complicate the solution\cite{Bena}. However, we emphasize
that on general ground, any spin dependent interaction, such as the
Rashba interaction, can lead to spin oscillations. In the single
electron phase coherent propagation, the spin transport is determined
by the quantum interference between partial waves, the
magnetoconductance due to a Rashba interaction is also expected to
differ from the classical regime although the details need to be
worked out.

Now, we return to the simple situation discussed in this paper and
address the question of experimental feasibility.  To realize coherent
ballistic precession, the length of the device must be long enough to
allow for a full spin revolution in the double-barrier region but
short enough for electrons to propagate coherently and ballistically.
Magnetic fields of 1 Tesla set the former limit to lengths larger than
$1 \mu m$. On the other hand, mean free paths in excess of $10 \mu m$
have been achieved in clean semiconductor heterojunctions.
Non-zero temperature will lead to decoherence due to phonons interaction. Moreover, because the
interference phenomena depend crucially on the phase accumulated
between the two barriers, the phase accumulated by the two {\em
  longitudinal} modes in the absence of a magnetic field should be
much smaller than $2\pi$.  This also sets a limit for the temperature
operation range and spin chemical potential to be less than $\hbar L/
v_F \approx 10 K$. Above this energy scale, thermal broadening and
energy dependent tunneling elements would obliterate the quantum
interference.   Another experimental challenge is the measurement of
the spin currents. Here we suggest that spin currents be obtained
through measuring the spin polarization of the contacts. Indeed, in
the absence of magnetic forces, the spin being conserved, the flow of
spin into the contacts can be related to the time-dependent spin
polarization of the contact ($I^\nu= \partial_{t} S^\nu$).  In
conclusion, we have analyzed spin transport in double barrier system
in the phase coherent and incoherent regimes. We showed that for
coherent propagation, the standard Bolztmann equations for spin
transport do not capture the rich behavior which can however be understood in
terms of quantum interference between multiple wave reflections. While completing this paper, we became aware of the work of Mireles {\em et al.} \cite{Mireles} on spin injection in the quantum coherent regime where similar conclusion were reached.  

\begin{acknowledgments}
  
This work was supported by the 
NSF through grant DMR-9985255, and the Sloan and Packard foundations.
\par
\end{acknowledgments}


\begin{thebibliography}{99}

\bibitem{Datta}  S. Datta and B. Das, Appl. Phys. Lett. {\bf 56}, 665 (1990).
\bibitem{Awschalom} D.D. Awschalom, D. Loss and D. Samarth, {\em Semiconductor Spintronics and Quantum Computation}, Springer Berlin, (2002) and references therein.
\bibitem{Kikkawa0} J.M. Kikkawa, I.P. Smorchkova, N. Samarth and D.D. Awschalom, Science {\bf 277}, 1284 (1997)
\bibitem{Kikkawa1} J.M. Kikkawa and D.D. Awschalom, Phys. Rev. Lett. {\bf 80},
4313 (1998).
\bibitem{Valet} T. Valet and A. Fert, Phys. Rev. B {\bf 48} 7099 (1993).
\bibitem{Johnson1} M. Johnson, Phys. Rev. B {\bf 58}, 9635, (1998)
\bibitem{Johnson} M. Johnson, Science {\bf 260}, 320 (1993).
\bibitem{Jedema} F.J. Jedema, A.T. Filip, B.J. van Wess, Nature  {\bf 410}, 345 (2001).
\bibitem{DattaBook} S. Datta, {\em Electronic Transport in Mesoscopic Systems},
Cambridge University Press, Cambridge, (1995).
\bibitem{ImryBook} I. Yoseph, {\em Introduction To Mesoscopics Physics}, Oxford University Press (1997). 
\bibitem{Kikkawa2} J.M. Kikkawa and D.D. Awschalom, Science, {\bf 287},
473 (2000); R. K. Kawakami et. al., Science, {\bf 294}, 131 (2001).
\bibitem{Buttiker1986} M. B\"uttiker, Phys. Rev. {\bf B33}, 3020 (1986);
IBM J. Res. Dev. {\bf 32}, 63 (1988).
\bibitem{Knittle} I. Knittle, F. Gagel and M. Schreiber, Phys. Rev. B.
{\bf 60}, 916 (1999).
\bibitem{Tang} H.X. Tang et al. in { \em Semiconductors Spintronics and Quantum Computation}, edited by D.D. Awschalom, D. Loss, N. Samarth, Springer Berlin,(2002).
\bibitem{Malajovich} I. Malajovich, J.M. Kikkawa, D.D. Awschalom, J.J. Berr, N. Samarth, Phys. Rev. Lett. {\bf 84}, 1015 (2000). 
\bibitem{Bena} C. Bena and L. Balents Phys. Rev. B {\bf 65}, 115108 (2002).
\bibitem{Mireles} F. Mireles and G. Kirczenow, cond-mat/0210391
\end{thebibliography}
\end{document}